\documentstyle[prd,aps,floats,tighten,epsfig,graphicx,color]{revtex}

\newcommand{\beq}{\begin{equation}}
\newcommand{\eeq}{\end{equation}}
\newcommand{\bea}{\begin{eqnarray}}
\newcommand{\ena}{\end{eqnarray}}
\newcommand{\etal}{{\it et al.}}
\newcommand{\ie}{{\it i.e.}}
\newcommand{\lsim}{\mathrel{\mathop{\kern 0pt \rlap
{\raise.2ex\hbox{$<$}}}
\lower.9ex\hbox{\kern-.190em $\sim$}}}
\newcommand{\gsim}{\mathrel{\mathop{\kern 0pt \rlap
{\raise.2ex\hbox{$>$}}}
\lower.9ex\hbox{\kern-.190em $\sim$}}}

\newcommand{\hepph}[1]{{\tt hep-ph/#1}}
\newcommand{\astroph}[1]{{\tt astro-ph/#1}}

\newcommand{\plb}[3]{Phys.\ Lett.\ B\ {\bf #1}, #3 (#2)}

\renewcommand{\apj}[3]{Astrophys.\ J.\ {\bf #1}, #3 (#2)}

\newcommand{\pr}[3]{Phys.\ Rev.\ {\bf #1}, #3 (#2)}
\renewcommand{\prl}[3]{Phys.\ Rev.\ Lett. {\bf #1}, #3 (#2)}
\renewcommand{\prd}[3]{Phys.\ Rev.\ D\ {\bf #1}, #3 (#2)}
\renewcommand{\plb}[3]{Phys.\ Lett.\ B\ {\bf #1}, #3 (#2)}

\newcommand{\href}[2]{#1}

\newcommand{\msol}{\mbox{$M_{\odot}$}}

\newcommand{\omegaL}{\mbox{$\Omega_{\Lambda}$}}

\definecolor{cyan}{cmyk}{1.,0.,0.,0.5}
\definecolor{magenta}{cmyk}{0.,1.,0.,0.5}
\definecolor{verdatre}{cmyk}{0.5,0.,0.5,0.5}
\definecolor{yellow}{cmyk}{0.,0.,0.2,0.0}
\definecolor{rouge}{cmyk}{0.,0.4,0.6,0.0}
\definecolor{orange}{cmyk}{0.,0.5,0.5,0.}
\definecolor{violet}{rgb}{0.5,0.,0.5}

\begin{document}

\draft
\title{Cosmological constraints on quintessential halos}
\vskip 1.cm
\author{
Alexandre Arbey$^{\rm a,b}$
\footnote{E--mail: arbey@lapp.in2p3.fr, lesgourg@lapp.in2p3.fr,
salati@lapp.in2p3.fr},
Julien Lesgourgues$^{\rm c,a}$ and Pierre Salati$^{\rm a,b}$
}
\vskip 0.5cm
\address{
\begin{flushleft}
a) Laboratoire de Physique Th\'eorique LAPTH, B.P.~110, F-74941
Annecy-le-Vieux Cedex, France.\\
b) Universit\'e de Savoie, B.P.~1104, F-73011 Chamb\'ery Cedex,
France.\\
c) Theoretical Physics Division, CERN, CH-1211 Gen\`eve 23, Switzerland. 
\end{flushleft}
}
\maketitle
\vskip 0.5cm
\centerline{6 February 2001}

\vskip 0.5cm
\begin{abstract}
A complex scalar field
has recently been suggested to bind galaxies and flatten the
rotation curves of spirals. Its cosmological behavior is
thoroughly investigated here.
Such a field is shown to be a potential candidate for the
cosmological dark matter that fills up a fraction
$\Omega_{\rm \, cdm} \sim 0.3$ of the Universe.
However, problems arise when the limits from galactic dynamics
and some cosmological constraints are taken simultaneously
into account.
A free complex field, associated to a very small mass 
$m \sim 10^{-23}$ eV, has a correct cosmological behavior in the
early Universe, but behaves today mostly as a real axion, 
with a problematic value of its conserved quantum number. 
On the other hand,
an interacting field with quartic coupling $\lambda \sim 0.1$
has a more realistic mass $m \sim 1$ eV and carries
a quantum number close to the photon number density.
Unlike a free field, it would be spinning today in the complex plane
-- like the so-called ``spintessence''. Unfortunately,
the cosmological evolution of such field in the early Universe
is hardly compatible with constraints from nucleosynthesis and
structure formation.
\end{abstract}
%
\vskip 1.cm

\section{Introduction}
\label{sec:introduction}

\vskip 0.1cm 
The observations of the Cosmic Microwave Background (CMB)
anisotropies \cite{boomerang}, combined either with the
determination of the relation
between the distance of luminosity and the redshift of supernovae
SNeIa \cite{supernovae_omegaL}, or with the large scale structure
(LSS) information from galaxy and cluster surveys \cite{2dF}, give
independent evidence for a dark matter density in the range
$\Omega_{\rm \, cdm} \, h^{2} = 0.13 \pm 0.05$ \cite{boomerang}, 
to be compared to a baryon density of
$\Omega_{\rm b} \, h^{2} = 0.019 \pm 0.002$ as indicated by
nucleosynthesis \cite{nucleosynthesis} and the relative heights
of the first acoustic peaks in the CMB data. The nature of that
component is still unresolved insofar. The favorite candidate for
the non--baryonic dark matter is a weakly--interacting massive particle
(WIMP). The so--called neutralino naturally arises in the framework of
supersymmetric theories. Depending on the numerous parameters of the model,
its relic abundance $\Omega_{\rm \, cdm}$ falls in the ballpark
of the measured value. New experimental techniques have been developed
in the past decade to detect these evading species.
However, detailed numerical simulations have recently pointed to a few
problems related to the extreme weakness with which that form of matter
interacts. Neutralinos tend naturally to collapse in numerous and highly
packed clumps \cite{moore} that are not seen -- see however \cite{navarro}.
The halo of the Milky Way should contain $\sim$ half a thousand satellites
with mass in excess of $10^{8}$ $\msol$ while a dozen only of
dwarf--spheroidals are seen.
The clumps would also heat and eventually shred the galactic ridge. More
generally, this process would lead to the destruction of the disks of spirals.
A neutralino cusp would form at their centers. This is not supported
by the rotation curves of low--surface--brightness galaxies that
indicate on the contrary the presence of a core with constant density.
Finally, two--body interactions with halo neutralinos and its associated
dynamical friction would rapidly slow down the otherwise observed spinning
bars at the center of spirals like M31.

\vskip 0.1cm
New candidates for the astronomical dark matter are therefore under
scrutiny such as warm dark matter \cite{wdm}, particles with self
interactions \cite{spergel}, or non--thermally produced WIMPs \cite{lin}.
An exciting possibility would be to have a common explanation for both
the dark energy $\omegaL$ and the dark matter $\Omega_{\rm \, cdm}$
components of the Universe. Before trying to reach such an ambitious 
goal\footnote{a possible
direction for using a quintessence field as dark matter was proposed
in \cite{sahni}.},
one could explore the relevance of scalar fields to the cosmological
and galactic dark matter puzzles, as was done for dark energy with the
so-called ``quintessence'' models \cite{quintessence}. The archetypal
example of quintessence is a neutral scalar field $\varphi$ with
Lagrangian density
\beq
{\cal L} \; = \; \frac{1}{2} \, g^{\, \mu \nu} \,
\partial_{\mu}  \varphi \, \partial_{\nu}  \varphi
\, - \, V \left( \varphi \right) \;\; .
\label{scalar_neutral}
\eeq
Should the field be homogeneous, its cosmological energy density
would be expressed as
\beq
\rho_{\varphi} \equiv T^{0}_{\; 0} \; = \;
{\displaystyle \frac{\dot{\varphi}^{2}}{2}} \, + \,
V \left( \varphi \right) \;\; ,
\label{mass_density}
\eeq
whereas the pressure would obtain from
$T_{i j} \equiv - g_{\, i j} \, P$ so that
\beq
P_{\varphi} \; = \; {\displaystyle \frac{\dot{\varphi}^{2}}{2}}
\, - \, V \left( \varphi \right)
\;\; .
\label{pressure}
\eeq If the kinetic term ${\dot{\varphi}^{2}}/{2}$ is small with
respect to the contribution from the potential $V \left( \varphi
\right)$, the equation of state can match the condition for driving
accelerated expansion in the Universe, $\omega_{Q} \equiv P_{\varphi}
/ \rho_{\varphi} < - 1/3$. Instead, in order to behave as
dark matter today, the field should be pressureless:
$|P_{\varphi}| \ll \rho_{\varphi}$.  So, the kinetic and potential
energies should cancel out in Eq.~(\ref{pressure}), a condition
automatically fulfilled by a quickly oscillating scalar field
averaged over one period of oscillation. This well--known setup is
that of the cosmological axion. It requires a quadratic scalar
potential, so that the kinetic and potential energies both redshift
as $\varphi^2 \propto a^{-3}$ with the Universe expansion and cancel
out at any time during the field dominated stage, which is then
equivalent to the usual matter dominated one.

\vskip 0.1cm
Axions -- or more generally, bosonic dark matter -- were revived
recently due to the undergoing CDM crisis. For instance, it was
noticed in \cite{hu} that stucture formation on small scales can be
forbidden by quantum mechanics, for wavelengths smaller than the
Compton wavelength -- i.e., the minimal spreading of an individual boson
wave function. The latter matches the scale of galactic substructures
for an ultra--light mass of order $m \sim 10^{-22}$~eV. Alternatively,
one may introduce a self--coupling term \cite{peebles,goodman}. As we
have seen, the existence of a matter--like dominated stage requires
that the contribution of non--quadratic terms to the potential energy
remains subdominant. Nevertheless, a self--coupling would modify the
field behavior in the early Universe, as well as its clustering
properties today in regions where the field is overdense -- exactly
like for boson stars, which are crucially affected by the presence of
a self--coupling \cite{colpi}. The self--coupling is also relevant
for the issue of field clumps stability, and can explain why dwarf
and low--surface--brightness galaxies have cores with finite
density \cite{riotto}.

\vskip 0.1cm
A remarkable feature with bosonic dark matter is the possibility to
form Bose condensates, i.e., large domains where the field is coherent
in phase and is in equilibrium inside its own gravitational potential
-- like boson stars -- or in that of an external baryonic matter
distribution. This opens the possibility to have a very simple and
elegant model for galactic halos, in which rotation curves would follow
from simple equations -- essentially the Klein--Gordon wave equation,
which governs classical scalar fields as well as Bose condensates. This
situation strongly differs from the more conventional picture of a gas
of individual particles -- fermions or heavy bosons -- for which
gravitational clustering does not lead to universal density profiles
and where the shape of galactic halos can only be studied through
technically difficult $N$--body simulations.

\vskip 0.1cm
The formation and stability of such condensates is a complicated issue
-- see e.g. \cite{tkachev,seidel,khle} -- even when the field is complex
and has a global charge -- not to be understood as an electric charge,
but as a conserved number of quanta like the baryon or lepton number.
For instance, a large condensate can be unstable under fragmentation
into smaller clumps.
For a real scalar field with no conserved charge, the issue of stability
is even more subtle since the field can self--annihilate, especially
when the condensate core density exceeds a critical value \cite{tkachev}.
This property can improve the agreement with observations \cite{riotto},
since the coupling constant will tune the upper limit on the density of
dark matter cusps at the center of galaxies. However, such a positive
feature is far from excluding models with a conserved charge. In fact,
the issue of Bose condensation on galactic scales -- in an expanding
Universe -- has never been studied in details. The result would depend
very much on the scalar potential, and it is difficult to guess what
would be the maximal core density today.

\vskip 0.1cm
In this work, we will focus on the scenario with a conserved charge,
and assume that dark matter consists in a complex scalar field with
a quasi--homogeneous density in the early Universe, producing later
galactic halos through Bose--condensation. The Lagrangian density
reads
\beq
{\cal L} \; = \;
g^{\mu \nu} \, \partial_{\mu} \phi^{\dagger} \, \partial_{\nu} \phi
\; - \; V \left( \phi \right) \, .
\eeq
Throughout this analysis, we will focus on the potential
\beq
V \; = \; m^{2} \phi^{\dagger} \phi \, + \,
\lambda \, \left\{ \phi^{\dagger} \phi \right\}^{2} \;\; .
\eeq
As a prologue to the study of density perturbations, we will follow
the evolution of the homogeneous cosmological background of this field,
taking into account the constraints on the scalar potential coming
from galactic halos.
In a previous work \cite{AJP_1}, we did a detailed comparison of such
halos with universal galactic rotation curves \cite{persic_salucci_stel},
for a massive complex scalar field. We recall the salient features of
this analysis in section~\ref{sec:galactic} and discuss how the corresponding
constraint on the mass $m$ is modified when a self--interaction coupling
$\lambda$ is introduced. In the subsequent section~\ref{sec:cosmological},
we study how this scenario can be implemented cosmologically. Neglecting
the spatial variations of $\phi$, we are led to the cosmological density
\beq
\rho_{\phi} \; = \; \dot{\phi}^{\dagger} \dot{\phi} \, + \,
V \left( \phi \right)
\label{mass_density_spin}
\eeq
and pressure
\beq
P_{\phi} \; = \; \dot{\phi}^{\dagger} \dot{\phi} \, - \,
V \left( \phi \right) \;\; .
\label{pressure_spin}
\eeq
Beside the issue of charge conservation, the case for a complex scalar
field is somewhat richer than that of a real (neutral) scalar field.
In one limit, the complex field can behave as an effective real one,
similar to the usual axion. On the other hand, it can be spinning in
the complex plane with slowly--varying modulus, like in the so--called
spintessence \cite{spintessence} scenario. This depends on the dominant
term in the kinetic energy, which can be either radial and oscillating,
or orthoradial and slowly varying. As a result, during the field dominated
era, the spintessence would have a continuously vanishing pressure,
while the axion pressure would oscillate between $+\rho_{\phi}$ and
$-\rho_{\phi}$. Finally, in section~\ref{sec:discussion}, we discuss
our results and feature the problems that plague the axion--spintessence
dark matter scenario. We will suggest some future directions worth being
investigated.

\section{Galactic halos}
\label{sec:galactic}
\vskip 0.1cm
We are interested in galactic halos consisting in self--gravitating
scalar field configurations -- which can be seen as Bose--Einstein
condensates spanning over very large scales. Since the typical
velocities observed in galaxies do not exceed a few hundreds of
km.s$^{-1}$, it is enough to study the quasi--Newtonian limit where
the deviations from the Minkowski metric
$\eta_{\mu \nu} = {\rm diag}(1,-1,-1,-1)$ are accounted for by the
vanishingly small perturbation $h_{\mu \nu}$. Inside galaxies, the
latter is of order the gravitational potential
\beq
h_{\mu \nu} \sim \Phi \sim V_{C}^{2} \;\; ,
\eeq
where $V_{C}$ is the rotation velocity -- in the case of spirals --
and where $\sqrt{2} \, V_{C}$ is the escape velocity from the system.
In the harmonic coordinate gauge where it satisfies the condition
\beq
\partial_{\alpha} h^{\alpha}_{\; \mu} \, - \, \frac{1}{2}
\partial_{\mu} h^{\alpha}_{\; \alpha} \; = \; 0 \;\; ,
\eeq
the perturbation $h_{\mu \nu}$ is related to the source tensor
\beq
S_{\mu \nu} \; = \; T_{\mu \nu} \, - \,
\frac{1}{2} \, g_{\mu \nu} \, T^{\alpha}_{\; \alpha}
\eeq
through
\beq
\Box h_{\mu \nu} \; = \;
- \, 16 \pi \, {\cal G} \; S_{\mu \nu} \;\; .
\eeq
The Poisson equation reads like
\beq
\Delta \Phi \; = \; 8 \pi \, {\cal G} \; S_{00} \;\; ,
\eeq
where $\Phi = h_{00} / 2$ is the Newtonian potential. For pressureless
matter, $2~S_{00} = T_{00} = \rho$. On the other hand, for the complex
scalar field, the gravitational potential is sourced by the effective
mass density
\beq
{\displaystyle \frac{\rho^{\rm eff}_{\phi}}{2}} \equiv S_{00} \; = \;
\left\{ 2 \, \dot{\phi}^{\dagger} \dot{\phi} \, - \,
V \left( \phi \right) \right\}
\label{mass_density_spin_galactic_1}
\eeq
which is a priori different from its cosmological counterpart
(\ref{mass_density_spin}). So, inside a galactic halo, the
gravitational potential is given by
\beq
\Delta \Phi \; = \; 4 \pi \, {\cal G} \,
\left( \rho^{\rm eff}_{\phi} \, + \, \rho_{\rm b} \right) \;\; ,
\label{newton}
\eeq
where $\rho_{\rm b}$ is the distribution of baryonic matter forming
the various galactic components -- stellar disk, bar, bulge... In first
approximation, the galaxy can be seen as spherically symmetric. In
that case, one shows \cite{tdlee} that all stable field configurations
must be in the form
\beq
\phi(r,t) \; = \;
{\displaystyle \frac{\sigma(r)}{\sqrt{2}}} \,
e^{\displaystyle i \, \omega t} 
\eeq
where the amplitude $\sigma$ depends only on the radius $r$.
Then, the effective field density reads like
\beq
\rho^{\rm eff}_{\phi} \; = \; \left\{ 2 \, \omega^{2} \sigma^{2} \, - \,
m^{2} \sigma^{2} \, - \, \frac{1}{2} \lambda \; \sigma^{4} \right\}
\;\; .
\label{mass_density_spin_galactic_2}
\eeq
The radial distribution of the field $\sigma(r)$ and the gravitational
potential $\Phi(r)$ are given by a system of two coupled equations:
the Poisson equation~(\ref{newton}) and the Klein--Gordon equation.
The latter may be expressed as
\beq
e^{\displaystyle - 2 v} \, \left\{ \sigma'' \, + \,
\left( u' + v' + \frac{2}{r} \right) \, \sigma' \right\} \; + \;
\omega^{2} e^{\displaystyle - 2 u} \, \sigma \; - \;
m^{2} \, \sigma \; - \; \lambda \, \sigma^{3} \; = \; 0
\label{Klein_Gordon_2}
\eeq
in the isotropic metric where
\beq
d \tau^{2} \; = \;
e^{\displaystyle 2 u} \, dt^{2} \, - \,
e^{\displaystyle 2 v} \, \left\{
dr^{2} + r^{2} d \theta^{2} + r^{2} \sin^{2} \theta \, d \varphi^{2}
\right\} \;\; .
\label{metric_isotropic}
\eeq
The prime denotes the derivative with respect to the radius $r$.
The Newtonian approximation corresponds to
\beq
u \; \simeq \; - \, v \; \simeq \; \Phi \;\; .
\eeq
Relation~(\ref{Klein_Gordon_2}) simplifies into
\beq
\Delta \sigma \; + \;
\left( 1 \, - \, 4 \Phi \right) \, \omega^{2} \, \sigma
\; - \;
\left( 1 \, - \, 2 \Phi \right)
\left( m^{2} \, \sigma \; + \; \lambda \, \sigma^{3} \right)
\; = \; 0 \;\; .
\label{KG}
\eeq
For each value of the parameters ($m$, $\lambda$, $\omega$) and
a given baryon distribution, these equations form an eigenvalue
problem with discrete solutions, labelled either by the central
value $\sigma_0>0$ or by the number of nodes $n$ in which
$\sigma(r)=0$. The lowest energy state -- which is not identically
null due to charge conservation -- has $n=0$. The self--consistency
of the Newtonian limit requires $|\Phi| \ll 1$. Such solutions
exist only for
\beq
0 < (m^2 - \omega^2) \ll m^2 \;\; .
\eeq

\subsection{Free field}
\label{subsec:free_field}
\vskip 0.1cm
In \cite{AJP_1}, we solved these equations for $\lambda = 0$.
We found that halos consisting in the fundamental configuration
of a free scalar field fit perfectly well the universal rotation
curves of low--luminosity spiral galaxies \cite{persic_salucci_stel}.
These data has three advantages for our purpose:
the robustness of the points and error bars -- obtained by averaging
over many galaxies --,
the good determination of the baryon distribution -- solely a stellar
disk with exponential luminosity profile --
and the low baryon contribution which justifies the approximation
of spherical symmetry.

\vskip 0.1cm
With a quadratic potential, the size of the halo is given by
\beq
l \; \sim \; \sqrt{\frac{M_P}{\sigma_{0}}} \,
\frac{\hbar}{m \, c} \;\; ,
\label{size1}
\eeq 
where we neglected the dependence on the baryon density. If the central
field value $\sigma_{0}$ is significantly smaller than the Planck mass,
the coherence length of the condensate exceeds the Compton wavelength
of an individual particle --
$l_{\rm compton} = \hbar / (m \, c)$ -- but it is clear that only an
ultra--light scalar field can condensate on distances of order 10 kpc.
The typical orbiting velocity in such a halo is given by
$v/c \sim \sqrt{\sigma_{0}/M_P}$. Therefore, requiring
$v \sim 100$~km.s$^{-1}$ and $l \sim 10$~kpc fixes $\sigma_{0}$ around
$10^{-6}~M_P$ and $m$ around $10^{-23}$~eV, as confirmed by a detailed
fitting to the data \cite{AJP_1}.

\vskip 0.1cm
Since the distribution of such halos only depends on the free parameters
$\sigma_{0}$ and $m$ -- where we impose a unique value of $m$ for all
galaxies -- we believe that their success in reproducing universal rotation
curves is a significant argument in favor of this model. On the other
hand, the existence of such a low mass, even if not strictly forbidden
by fundamental principles, is very unlikely due to unavoidable radiative
corrections. This could motivate a systematic investigation of other
potentials for the scalar field. The next level of complexity would
consist in adding a quartic self--coupling.

\subsection{Quartic self--coupling $\lambda$}
\label{subsec:self_coupling}
\vskip 0.1cm
As is well known for boson stars -- which are exactly identical to our
halos in the absence of a baryon component -- the inclusion of a quartic
term drastically modifies the mass $m$ and the extension of the condensate,
even when it contributes to a negligeable fraction of the central energy
density \cite{colpi,liddle}. This is so because $\lambda \, \sigma^{2}$ can
be very small with respect to $m^{2} \simeq \omega^{2}$ and yet comparable
to the difference $(m^{2} - \omega^{2})$ that appears in the Klein--Gordon
equation. In the limit where
$\Lambda \equiv \lambda \, / (4 \pi \, {\cal G} \, m^{2}) \gg 1$ and in the
absence of a baryon population, we can even give an exact analytic solution
for the field and for the orbiting velocity of test particles:
\begin{eqnarray}
\sigma(r) &=& \sigma_{0} \left\{ {\displaystyle
\frac{\sin \left( m \sqrt{2/\Lambda} \, r \right)}{m \sqrt{2/\Lambda} \, r}}
\right\}^{1/2} \;\; , \label{field_behavior} \\
{\displaystyle \frac{v(r)}{c}} &=& r \, \Phi'(r) \; = \;
2 \pi ~\Lambda ~{\displaystyle \frac{\sigma_{0}^{2}}{M_P^{2}}}
\left\{ {\displaystyle
\frac{\sin \left( m \sqrt{2/\Lambda} \, r \right)}{m \sqrt{2/\Lambda} \, r}}
\, - \, \cos \left( m \sqrt{2/\Lambda} \, r \right) \right\} \;\; ,
\end{eqnarray}
with the requirement that 
\beq
\Lambda^{-1} \ll \frac{\sigma_{0}}{M_P} \ll
\Lambda^{-1/2} \;\; .
\label{condition}
\eeq
Because ${\sigma_{0}}/{M_P} \sim {(v/c)}/{\sqrt{\Lambda}}$, the second
inequality follows from the Newtonian self--consistency condition
$|\Phi| \sim v^{2} \ll 1$. The first inequality translates into
$\Lambda \gg \left( c/v \right)^{2}$. It implies that all the field
spatial derivatives can be neglected in Eq.~(\ref{KG}) and sets the
maximal radius up to which the analytic solution is valid. This maximal
radius is at most equal to half a period so that
$r \leq \sqrt{\Lambda/2} \, (\pi / m)$. That bound is almost saturated for
$\Lambda \rightarrow \infty$. Note that $\omega$ does not appear in the
analytic solutions because it is only relevant at larger radii. However,
the Newtonian self--consistency condition imposes that
$1 - \omega^{2}/m^{2} \ll 1$.

\vskip 0.1cm
The field behavior~(\ref{field_behavior}) may be readily recovered
by neglecting the spatial derivatives of $\sigma$ in the Klein--Gordon
equation~(\ref{KG}) so that
\beq
\sigma^{2}(r) \; \simeq \; {\displaystyle \frac{m^{2}}{\lambda}} \,
\left\{ {\displaystyle \frac{\Omega^{2}}{B}} \, - \, 1 \right\}
\;\; ,
\eeq
where $B(r) = 1 \, + \, 2 \, \Phi(r)$ and $\Omega = \omega / m$. In the
Newtonian limit, the pressure~(\ref{pressure_spin}) reads like
\beq
P_{\phi} \equiv {\cal L} \; \simeq \;
{\displaystyle \frac{m^{4}}{4 \, \lambda}} \,
\left\{ {\displaystyle \frac{\Omega^{2}}{B}} \, - \, 1 \right\}^{2}
\;\; ,
\eeq
whilst the effective mass density~(\ref{mass_density_spin_galactic_2})
of the Bose--condensate is
\beq
\rho^{\rm eff}_{\phi} \; \simeq \; m^{2} \, \sigma^{2} \; \simeq \;
{\displaystyle \frac{m^{4}}{\lambda}} \,
\left\{ {\displaystyle \frac{\Omega^{2}}{B}} \, - \, 1 \right\}
\;\; .
\eeq
Both are related through the Lane--Emden polytropic equation
of state
\beq
P_{\phi} \; = \; K \, {\rho^{\rm eff}_{\phi}}^{\displaystyle 1 + 1/n} \;\; ,
\eeq
with $K = \lambda \, / (4 \, m^{4})$ while the polytropic index is $n=1$.
For such a value, the gravitating system -- in hydrostatic equilibrium --
is shown to have a constant core radius $r_{c} = \pi \, a$ where
\beq
a^{2} \; = \; {\displaystyle \frac{1}{8 \, \pi \, {\cal G}}} \;
{\displaystyle \frac{\lambda}{m^{4}}} \;\; .
\eeq
The field and density profiles are functions of the reduced radius
$z = r / a$
\beq
{\displaystyle \frac{\rho'_{\phi}(r)}{\rho'_{\phi}(0)}}
\; = \;
{\displaystyle \frac{\sigma^{2}(r)}{\sigma_{0}^{2}}} \; = \;
{\displaystyle \frac{\sin(z)}{z}} \;\; .
\eeq

\vskip 0.1cm
The most striking feature in the large $\Lambda$ limit is as follows:
although the quartic term remains subdominant in the energy density
-- Eq.~(\ref{condition}) implies that
$\lambda \sigma^{4} \ll m^{2} \sigma^{2}$ -- the typical size of the
system is very different from the free field case since now it reads like
\beq
l \sim \lambda^{1/2} \, M_P \; \hbar / (m^{2} c) \;\; .
\label{size2}
\eeq 
As the central field value does not appear in this expression,
different halo sizes would just result from different baryon contributions
to the density, which bring corrections to (\ref{size2}). The central
field value $\sigma_{0}$ still determines the rotation curve amplitude.
In the large $\Lambda$ limit and in the absence of baryons, the maximal
rotation speed is given exactly by
\beq
{\displaystyle \frac{v_{\rm max}^{2}}{c^{2}}} \; = \;
2.13~ \pi ~\Lambda \, {\displaystyle \frac{\sigma_{0}^{2}}{M_P^2}}
\qquad {\rm at} \qquad
r \; = \; 1.94 \, \Lambda^{1/2} \, / \, m \;\; .
\label{size3}
\eeq
In this paper, we will not extend our previous comparison with galactic
rotation curves \cite{AJP_1} to the case of a quartic self--coupling,
leaving this for a future publication. We will just use
relations~(\ref{size2}) and (\ref{size3}) in order to find a rough order
of magnitude estimate for $\lambda / m^{4}$ which has the best chance to
provide a good fit to the data. In the next section, this constraint will
be plugged into a cosmological dark matter scenario. By requiring that
the rotation velocity peaks around 200~km.s$^{-1}$ at a typical radius
of 10~kpc, we find
\beq
m \sim \lambda^{1/4} \; {\rm eV}
\qquad {\rm and} \qquad
\sigma_{0} \sim \Lambda^{-1/2} \; 10^{-3} \; M_P \;\; .
\label{relation_34}
\eeq
Taking for instance $\lambda$ in the range $[1 \, , \, 10^{-4}]$, we
obtain a mass of order 0.1 to 1 eV, {\ie}, a few orders of magnitude
larger than the expected neutrino masses. This gives
$\Lambda \sim 10^{52}$ and $\sigma_{0} \sim 0.1$~eV.

\section{Cosmological dark matter}
\label{sec:cosmological}
\vskip 0.1cm

\subsection{Matter--like behavior}

A homogeneous complex scalar field with a quadratic potential is
a perfect candidate for pressureless cold dark matter. Let us write
the Klein--Gordon and Friedman equations for the field configuration
\beq
\phi(t) \; = \; {\displaystyle \frac{\sigma(t)}{\sqrt{2}}} \,
e^{\displaystyle i \, \theta(t)} \;\; .
\eeq
This amounts to
\begin{eqnarray}
{\displaystyle \frac{d^2 \sigma}{d t^2}} +
\frac{3}{a} {\displaystyle \frac{da}{dt}}{\displaystyle \frac{d \sigma}{dt}}
+ m^2 \sigma + \lambda \sigma^3 - \omega^2 \sigma &=& 0 \;\; ,\\
{\displaystyle \frac{d \omega}{dt}} \sigma +
\frac{3}{a} {\displaystyle \frac{da}{dt}} \omega \sigma +
2 \omega {\displaystyle \frac{d \sigma}{dt}} &=& 0 \;\; ,\\ 
3 H^2 = 3 \left( {\displaystyle \frac{da}{a \, dt}} \right) ^2 =
8 \pi \, {\cal G} \; \left[ \, \rho_{\rm rad}^{\mbox{ }} \right.
&+&
\left.
\frac{1}{2} \left\{
\left( {\displaystyle \frac{d \sigma}{dt}} \right)^2 +
\omega^2 \sigma^2 + m^2 \sigma^2 + \frac{\lambda}{2} \sigma^4 \right\}
\right] \;\; 
\end{eqnarray}
where $\rho_{\rm rad}$ is the usual density of relativistic photons and
neutrinos and $\omega = d\theta / dt$. The second equation implies the
conservation of the charge per comoving volume $Q = \omega \sigma^2 a^3$.
Therefore, we can rewrite the first equation as
\begin{equation}
{\displaystyle \frac{d^2 \sigma}{d t^2}} +
\frac{3}{a} {\displaystyle \frac{da}{dt}}{\displaystyle \frac{d \sigma}{dt}}
+ m^2 \sigma + \lambda \sigma^3 -
{\displaystyle \frac{Q^2}{\sigma^3 a^6}} = 0 \;\; .
\end{equation}
It is most convenient to use the dimensionless time and field variables:
\beq
\tilde{t}=mt~, \qquad \tilde{H}=H/m, \qquad 
u = \sqrt{\frac{m}{Q}} a^{3/2} \sigma~.
\eeq
Then, the charge conservation implies $\omega \, u^2 = m$ and the
Klein--Gordon equation reads like
\beq
\ddot{u} + \left\{
1 - u^{-4} + \frac{\lambda Q}{a^3 m^3} u^2 - \frac{3}{2}
\left( \dot{\tilde{H}} + \frac{3}{2} \tilde{H}^2 \right)
\right\} u = 0 \;\;, 
\label{motionu}
\eeq
where the dot denotes a derivative with respect to $\tilde{t}$.
When $\lambda \neq 0$, we see that the self--coupling term always
dominates the mass term in the past, when $a \rightarrow 0$. However,
we will first study the opposite case when the quartic term is zero or
subdominant. If Eq.~(\ref{motionu}) is to be applied today, or in the
late stage of evolution of the Universe, we can also neglect the
contribution from the Universe expansion: today, one has
$H_0 \sim 10^{-61} M_P$, many orders of magnitude below the values of
$m$ discussed in section~\ref{sec:galactic}. So, Eq.~(\ref{motionu})
reduces to 
\beq
\label{eqforu}
\ddot{u} + u - u^{-3} = 0 \;\; ,
\eeq
and describes some periodic oscillations in a static potential
$V(u) \equiv (u^2 + u^{-2})/2$, with a minimum at $u_0=1$ corresponding
to $\omega=m$. The field density reads
\beq
\rho_{\phi} = m \, Q \, E_u \, a^{-3} \;\; ,
\label{densityFD}
\eeq
where $E_u \geq 1$ stands for the conserved energy
$(\dot{u}^2 + u^2 + u^{-2})/2$. We conclude that in the late evolution
of the Universe, as soon as $m$ becomes bigger than $H$ and
$\lambda^{1/2} \sigma$, the homogeneous scalar field behaves exactly as
a cosmological background of dark matter.

\vskip 0.1cm
Let us provide further intuition on the physical meaning of the
oscillations for the variable $u$.  If $E_u \simeq 1$ and $u$ is
almost stable around one, then the modulus slowly decreases as
$\sigma \propto a^{-3/2}$, while the phase velocity is constant with
$\omega=m$. The equation of state is that of pressureless matter:
\beq
w = \frac{P_{\phi}}{\rho_{\phi}} \sim \frac{H^2}{m^2} \ll 1 \;\; .
\eeq
Such a field, following a spiral trajectory, is a particular case of what
was recently called spintessence \cite{spintessence}. On the other hand,
when $E_u \gg 1$, $u$ and $\omega$ strongly oscillates, but the trajectory
of $u \, e^{\displaystyle i \theta(t)}$ is a fixed ellipse as can be seen
from the exact analytic solution to Eq.~(\ref{eqforu}):
\beq
u = \left\{ \left( E_u^2 - 1 \right)^{1/2}
\sin \left( 2 \tilde{t} + \alpha \right) + E_u \right\}^{1/2},
\qquad 
\theta \left( \tilde{t} = \pi \right) - \theta(0) =
\int_{\displaystyle 0}^{\displaystyle \pi/m} \! \! \! \! \! \omega(t) dt
= \int_{\displaystyle 0}^{\displaystyle \pi} u^{-2} \, d\tilde{t} = \pi
\;\; .
\label{ellipse}
\eeq
The integral of the phase over one period of oscillation obtains from 
Gradshteyn and Ryzhik, 3.661.4. So, the field $\phi$ follows an ellipse
which axes decrease as $a^{-3/2}$. In the large $E_u$ limit, the ellipse
is squeezed and reduces asymptotically to a line. Then, the complex field
is essentially similar to a real oscillating field, like in usual axionic
dark matter models. The pressure does not vanish identically, but quickly
oscillates between $+\rho_{\phi}$ and $-\rho_{\phi}$, with zero average over
one period $\Delta t = \pi/m$. This period is much smaller than $H^{-1}$: on
cosmological scales, the field still has the same effect as pressureless matter.

\subsection{radiation domination: the $\lambda=0$ case}

In the $\lambda=0$ case, we know that the mass should be of order
$m \sim 10^{-23}$ eV, ten orders of magnitude above the present value
of the Hubble parameter. It is easy to show that $H$ will start to dominate
over the mass at a redshift of order $6 \times 10^5$. Therefore, the
previous analytic solutions (\ref{densityFD}) and (\ref{ellipse}) only apply
to the end of the radiation dominated stage. During radiation domination,
the Klein--Gordon equation~(\ref{motionu}) simplifies into
\beq
\ddot{u} + u + \frac{3}{4} \tilde{H}^2 u - u^{-3} = 0 \;\; .
\eeq
When $H \gg m$ -- i.e., $\tilde{H} \gg 1$ -- the term $u$ can be neglected
in the above equation. We then obtain a simple non--linear equation for the
variable $v = a^{-1/2} u$
\beq
v'' - \frac{1}{v^3} = 0 \;\; ,
\label{KGl0}
\eeq
where the prime denotes derivation with respect to conformal time
$d \tilde{t} = a \, d \tau$. There is an exact solution:
\beq
\sigma = \sqrt{\frac{Q}{m}} \, \frac{v}{a} =
\left\{ C_1 + \frac{C_2}{a} + \frac{C_3}{a^2} \right\}^{1/2} \;\; .
\label{fieldRDl0}
\eeq
Only two of the three constants ($C_1$,  $C_2$, $C_3$) are independent since
\beq
C_1 C_3 = \frac{C_2^2}{4} + \frac{Q^2}{(a^2H)^2} \;\; .
\eeq
Remember that $a^2 H$ is also constant during radiation domination.
The free parameters ($C_1$, $C_2$) can be conveniently expressed as a
function of the initial conditions for the field at some initial time $t_i$:
\begin{eqnarray}
C_1 &=& \sigma_i^2 + \frac{2 \sigma_i (d\sigma / dt)_i}{H_i}
+ \frac{(d \sigma / dt)_i^2}{H_i^2} + \frac{\omega_i^2 \sigma_i^2}{H_i^2}
\;\; ,\\
C_2 &=& -2 a_i \left\{
\frac{ \sigma_i (d\sigma / dt)_i}{H_i} +
\frac{(d \sigma / dt)_i^2}{H_i^2} +
\frac{\omega_i^2 \sigma_i^2}{H_i^2}
\right\} \;\; .
\end{eqnarray}
These results show that $\sigma$ quickly stabilizes at a value $\sqrt{C_1}$.
During radiation domination, the field energy density reads like:
\beq
\rho_{\phi} \; = \;
{\displaystyle \frac{(a^2 H)^2}{2}}{\displaystyle \frac{C_3}{a^6}}
\, + \,
{\displaystyle \frac{m^2}{2}}
\left\{ C_1 + \frac{C_2}{a} + \frac{C_3}{a^2} \right\} \;\; .
\label{densityRDl0}
\eeq 
Interestingly, we see that after some time the energy can be dominated
by the contribution of the mass term, even if the latter can be neglected
in the Klein--Gordon equation.

\vskip 0.1cm
We now understand what the generic evolution looks like, starting from
an early time -- say, for instance, at the end of reheating -- at which
the field density is smaller than the radiation density:
first, $\rho_{\phi}$ decays as $a^{-6}$; then, it stabilizes around the
value $m^2 C_1/2$, and remains constant as long as $H>m$; finally,
when $H<m$, the density decays as that of pressureless matter, takes
over the radiation density, and drives the matter--like dominated stage.
Such a scenario requires a single constraint: the constant value of the
field density during radiation domination should be matched with the matter
density extrapolated from today back to the time $t_*$ at which $H=m$:
\beq
\frac{m^2}{2} \, C_1 \; = \; \Omega_{\rm \, cdm} \, \rho_c^0 \,
\left( \frac{a_0}{a_*} \right)^3
\qquad \Rightarrow \qquad
C_1 \sim 7 \times 10^{-4} \, M_P^2 \;\; .
\label{plateau}
\eeq
The scenario is enough constrained to reach another conclusion:
when the field density decays as matter, its dynamics is that of an
effective oscillating real scalar field and not that of spintessence.
Indeed, as soon as the density becomes constant during radiation
domination, one has
\beq
{\displaystyle \frac{(a^2 H)^2}{2}} \,
{\displaystyle \frac{C_3}{a^6}} \ll
{\displaystyle \frac{m^2}{2}} \, C_1
\qquad \Rightarrow \qquad
{\displaystyle \frac{Q^2}{a^6}} \ll m^2 C_1^2 \;\; .
\label{nospint}
\eeq
On the other hand, a crude matching between the two expressions for
the energy density (\ref{densityFD}) and (\ref{densityRDl0}) at the
time when $H=m$ gives:
\beq
\label{reallynospint}
m \, Q \, E_u \,a^{-3} \sim m^2 \, C_1
\qquad \Rightarrow \qquad
\frac{Q^2}{a^6} \, E_u^2 \sim m^2 \, C_1^2 \;\; .
\eeq
By combining Eq.~(\ref{nospint}) and (\ref{reallynospint}), we find
that $E_u^2 \gg 1$, which is the condition for the field trajectory
in the complex plane to be a squeezed ellipse of constant phase.

\vskip 0.1cm
So far, we have assumed that the radiation density was initially dominant,
in order to use the analytic solutions (\ref{fieldRDl0}) and
(\ref{densityRDl0}). Instead, if the field dominates initially -- a
situation that could be allowed before the time of Big Bang Nucleosynthesis
(BBN) -- the field density will nevertheless decay as $a^{-6}$. Indeed, if
$\rho_{\phi} \geq \rho_{\rm rad}$, then necessarily\footnote{Proof:
even if $\sigma$ was as large as $M_P$, $m^2 \sigma^2$ would be of order
$10^{-102} \, M_P^4$, which corresponds to the radiation density at redshift
$10^6$, well after BBN.} $\rho_{\phi} \gg m^2 \sigma^2$.
We conclude that $P_{\phi} = \rho_{\phi}$, and energy conservation implies
that $\rho_{\phi} \propto a^{-6}$. As soon as the radiation density takes
over, the previously described analytic solutions do apply.

\begin{figure}[h!]
\centering 
\leavevmode
\epsfxsize=16.5cm
\epsfbox{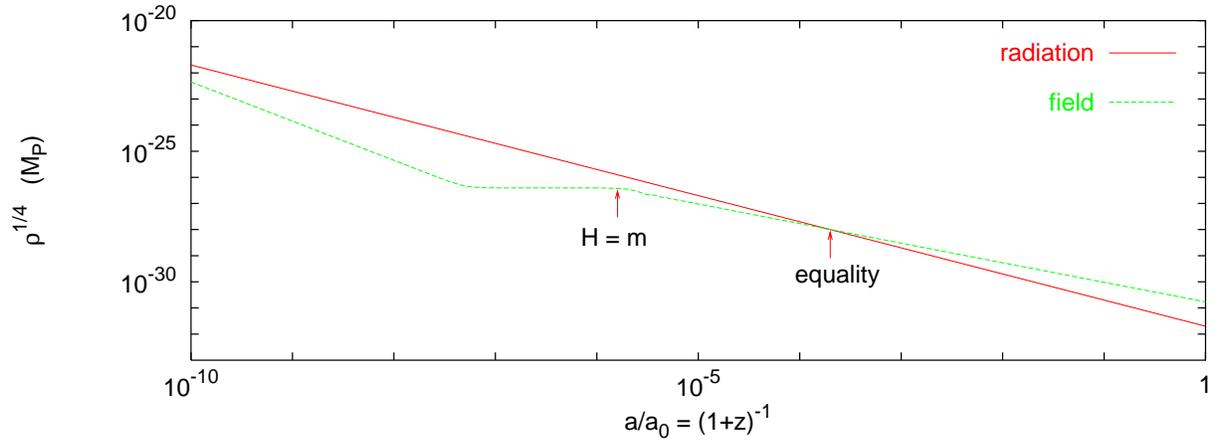}
\vspace{1cm}
\caption{Evolution of the field and matter densities, in the scenario 
with $\lambda=0$ and $m \sim 10^{-23}$~eV. The evolution starts at a
redshift $z = 10^{10}$ -- but we could have equally well started earlier.
The initial value of the field density can be chosen arbitrarily, below
or above that of radiation density. After decaying as $a^{-6}$, the field
density reaches a plateau which amplitude has been fixed according to
Eq.~(\ref{plateau}). This condition ensures a correct value of the density
today: $\rho_{\phi}=\Omega_{\rm \, cdm} \rho_{\rm c}^0$. At
$z=6.3 \times 10^5$, when $H=m$, the density starts to decrease as $a^{-3}$,
like for presureless matter, and takes over radiation at
$z \simeq 3 \times 10^3$.}
\label{fig:fig1}
\end{figure}

\begin{figure}[h!]
\centering 
\leavevmode
\epsfysize=7.cm
\epsfbox{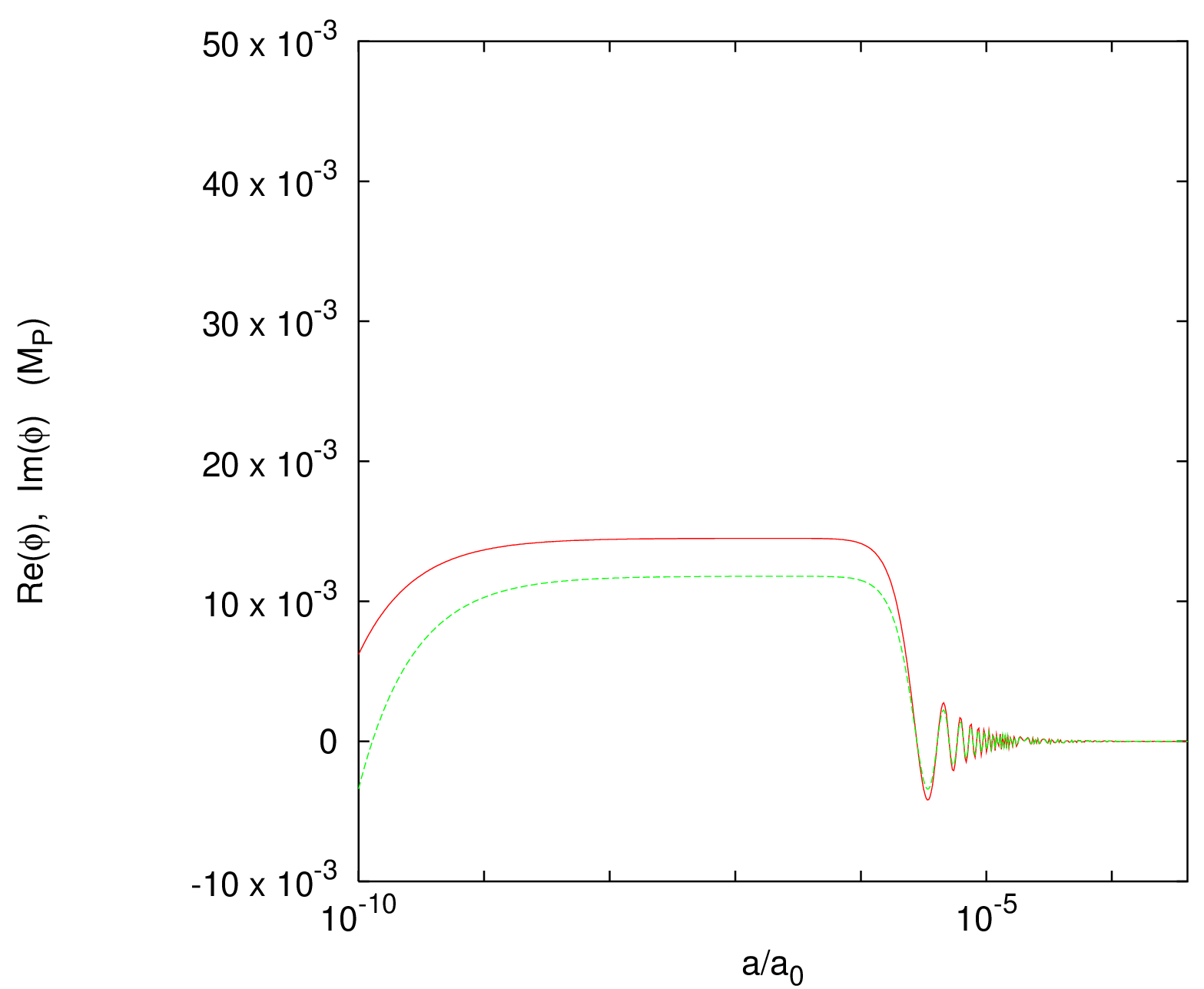}
\epsfysize=7.cm
\epsfbox{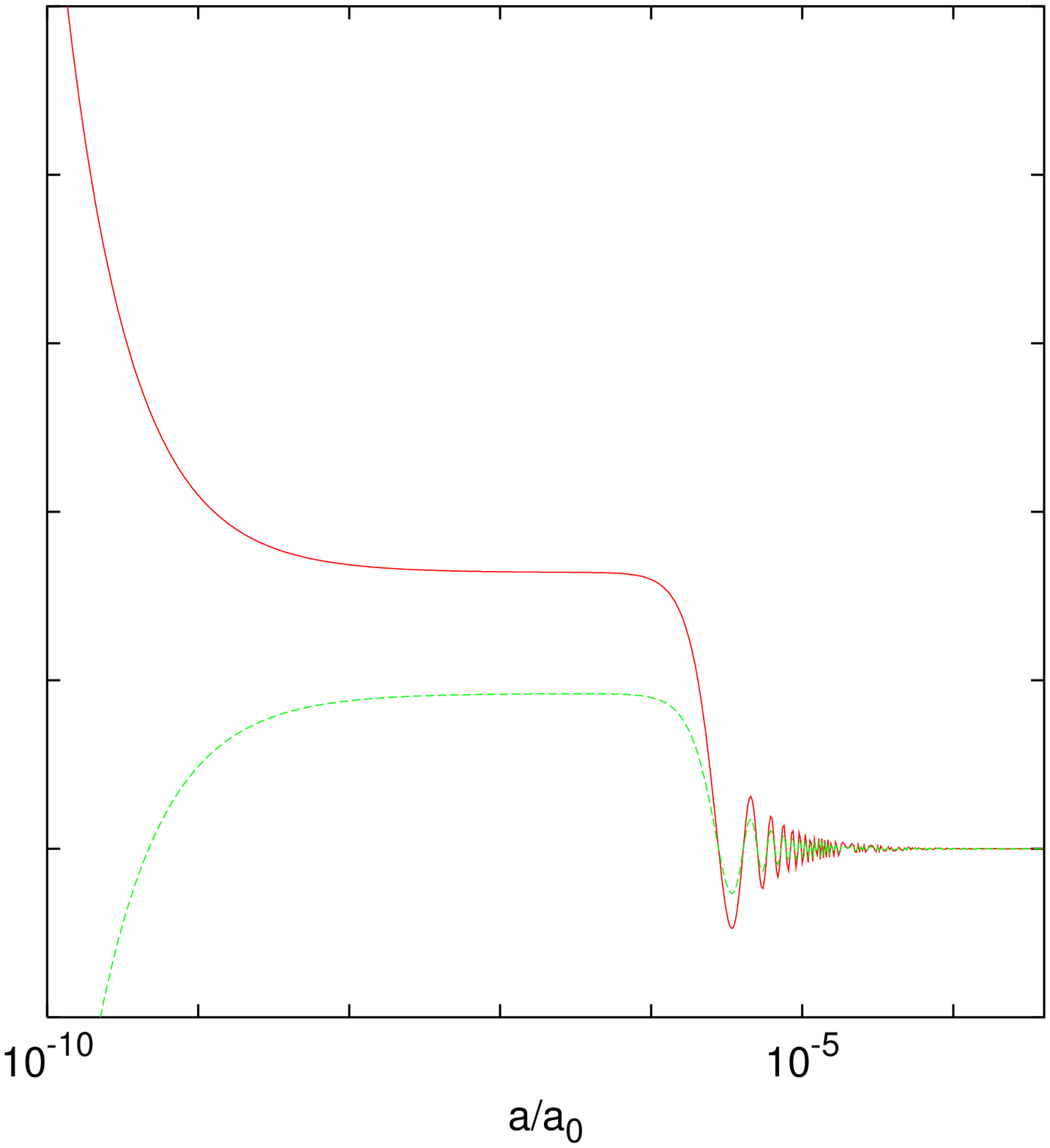}

\vspace{0.5cm}

\caption{Two possible evolutions of the field real and imaginary
parts, in the scenario with $\lambda=0$ and $m \sim 10^{-23}$~eV.
On the left panel, the initial field modulus has been chosen below
the equilibrium value $\sqrt{C_{1}}$ as determined in
Eq.~(\ref{plateau}).
The opposite situation is shown on the right panel. In both cases,
the field density decays like in Fig.~\ref{fig:fig1}. At redshift
$z=6.3 \times 10^5$, when $H=m$, the field starts to oscillate, but
its density decays smoothly as that of presureless matter. Because
the real and imaginary parts are exactly in phase, the field is
equivalent to a single real scalar field.}
\label{fig:fig2}
\end{figure}

\subsection{radiation domination: $\lambda \neq 0$}

If the field has got a quartic self--coupling, $\lambda \sigma^4$ must be
negligible with respect to $m^2 \sigma^2$ in the late Universe in order
to drive a matter--like dominated stage with $\rho_{\phi} \propto a^{-3}$.
However, a quartic self--coupling is likely to be cosmologically relevant
at early times, whenever the field modulus $\sigma$ well exceeds 
$\lambda^{-1/2} \, m$. In that case, the equation for $v$ -- see
relation (\ref{KGl0}) -- reads like
\beq
v'' - \frac{a''}{a} \, v + \frac{\lambda \, Q}{m^3} \, v^3 
- \frac{1}{v^3} = 0 \;\; .
\label{KGl}
\eeq
During radiation domination, $a''=0$ and $v$ is a periodic -- elliptic --
function, describing non--harmonic oscillations in the potential
$V(v) = \lambda \, Q \, v^4 / (4 \,m^3) \; + \; 1 / (2 \,v^2)$, with a
minimum at $v_0 = \left\{ m^3 / (\lambda Q) \right\}^{1/6}$. The period
of the oscillations -- expressed in conformal time -- is of order
$m (Q \lambda)^{-1/3}$. So, $\sigma$ performs damped oscillations
with a constant period $\Delta a \sim a^2 H (Q \lambda)^{-1/3}$ with
respect to the scale factor. If we furthermore define the conserved
pseudo--energy energy of $v$ by $E_v = v'^{2}/2 + V(v)$, we can express
the field density as
\beq
\rho_{\phi} = m \, Q \, \left\{ E_v - \frac{H a}{m} \, v \, v' +
\frac{H^2 a^2}{2 \, m^2} v^2 \right\} a^{-4} \;\; .
\label{densityl}
\eeq
Remember that $H a$ decays as $a^{-1}$ during radiation domination:
a priori, at early times, the field density performs damped oscillations
like the field modulus while at late times it decays smoothly, as for
radiation
\beq
\label{smoothdensityl}
\rho_{\phi} = m \, Q \, E_v a^{-4} \;\; .
\eeq
The transition between both behaviors takes place when
$H^2 a^2 v^2 \sim m^2 E_v$, where $v$ is evaluated at the maximum of one
oscillation: $v_{\rm max} \sim (4 m^3 E_v / \lambda Q)^{1/4}$. Inserting
this condition in Eq.~(\ref{densityl}), we find that the transition
between damped oscillations and smooth decay occurs when
$\rho_{\phi} \simeq H^4/ \lambda$. In practice, this implies that the
oscillatory behavior is generally irrelevant unless $\lambda$ is
fine--tuned to extremely small values -- for the ordinary radiation
component, the condition $\rho_{\rm rad} \gg H^4$ is already realized
at the end of inflation. So, Eq.~(\ref{smoothdensityl}) applies even in
the early Universe.

\vskip 0.1cm
Later on, the transition between radiation--like and matter--like
behaviors will be effective when the maximal value of $\sigma$ during
one oscillation, computed from Eq.(\ref{KGl}), will be comparable
to $\lambda^{-1/2} m$. This translates into
\beq
\label{transition}
\sigma_{\rm max} = \sqrt{\frac{Q}{m}}
\frac{v_{\rm max}}{a} = \frac{1}{a}
\left(
\frac{4 m Q E_v}{\lambda} \right)^{1/4} =
\frac{m}{\sqrt{\lambda}} \qquad \Rightarrow \qquad a =
\left( \frac{4 \lambda Q E_v}{m^3} \right)^{1/4} 
\qquad \Rightarrow \qquad \rho_{\phi} \sim
\frac{m^4}{\lambda} \simeq ({\rm eV})^4 \;\; .
\eeq
In the last equality we used the constraint from the size of galactic
halos. Since, on the other hand, $\rho_{\rm eq} \sim 0.55 \, ({\rm eV})^4$,
the transition to matter--like behavior occurs slightly before equality.
This means that at earlier times, when the field behaves like radiation,
its density should be fine--tuned in order to be close to the radiation density.

\begin{figure}[h!]
\centering 
\leavevmode
\epsfxsize=16.5cm
\epsfbox{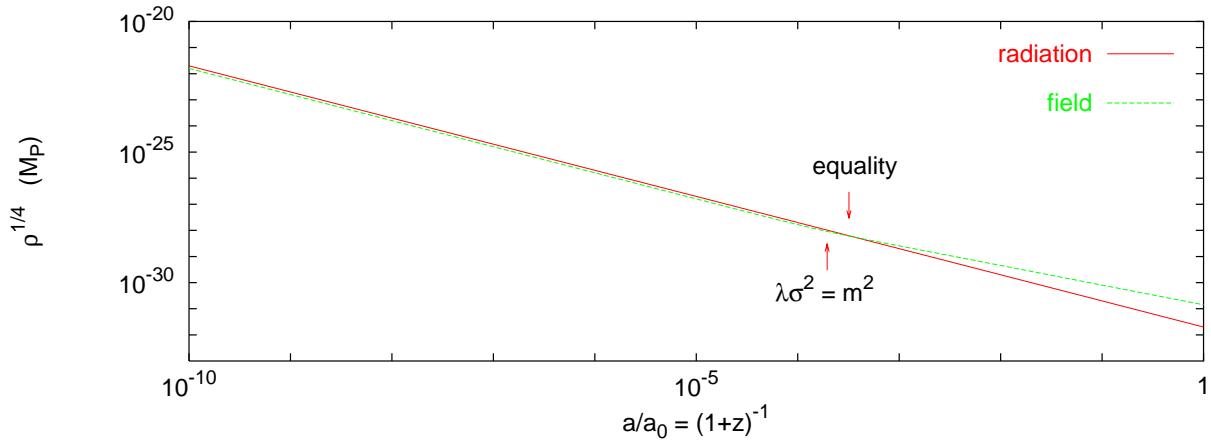}
\vspace{1cm}
\caption{Evolution of the field and matter densities, in the scenario
with $m \sim \lambda^{1/4}$~eV and arbitrary $\lambda$. The evolution
also starts at a redshift $z=10^{10}$. The field decays first as dark
radiation, and then as dark matter. The transition between these two
regimes is constrained by Eq.~(\ref{transition}) to take place immediately
before matter--radiation equality. In other words, in the early Universe,
the field density must be very close to that of radiation. The simulation
gives $\rho_{\phi} = 0.6 \, \rho_{\rm rad}$ in order to obtain a correct
value of $\rho_{\phi} = \rho_{\rm cdm}$ today
(such that $\Omega_{\rm cdm} \, h^2 = 0.13$).}
\label{fig:fig3}
\end{figure}

\vskip 0.1cm
This scenario is illustrated on Fig.~\ref{fig:fig3}. In order to obtain
the correct value of the field and radiation densities today, $\rho_{\phi}$
must be adjusted to $0.6 \, \rho_{\rm rad}$. This would correspond to an
effective number of extra neutrino species of 
$\Delta N_{\rm eff} = 5$ that is
not even allowed by BBN.

\section{Discussion}
\label{sec:discussion}

In section II, we recalled why a complex scalar field is an attractive 
candidate for dark matter in galactic halos; it can provide a
rather powerful explanation to the observed rotation curves, because
the radial distribution of a scalar field after Bose condensation 
is constrained by a wave equation (the Klein--Gordon equation);
this form of dark matter would be very smooth, and unlike a gas of
individual particles, it would not slow down the relative motion
of the baryonic components through dynamical friction.

\vskip 0.1cm
The results of section III show that the complex scalar field can also
play the role of a cosmological dark matter background, since after
various possible behaviors in the early Universe, which depend on the
exact shape of the potential, the field will decay like ordinary
pressureless matter as soon as the mass term dominates the potential
and the mass is bigger than the Hubble parameter.

\vskip 0.1cm
However, when the parameters of the scalar potential are estimated
from the size and mass of galactic halos, and plugged into the
cosmological evolution, some tension appears both for the free--field
model and for the one with a quartic self--coupling. Indeed, the redshift at
which the field starts to decay like matter, $\rho_{\phi} \propto
a^{-3}$, is completely fixed by these parameter values.

\vskip 0.1cm
In the case with $\lambda=0$, this transition occurs at a redshift
close to $z = 6 \times 10^5$. The comoving wavenumber of perturbations
entering into the Hubble radius at that time is $k=3$~Mpc$^{-1}$.  As
shown in Ref.\cite{hu}, this is of the same order of magnitude as the
Jeans length for the free field. So, larger wavelengths -- in
particular, those probed by the spectrum of CMB anisotropies and by
the linear matter power spectrum -- are expected to undergo the same
evolution as in an ordinary CDM model. On smaller wavelength, one
would still need to prove with numerical simulations that Bose
condensation can occur on the scale of galactic halos -- and eventually
also on slightly bigger and smaller scales. So far, this model still
sounds attractive, apart maybe from the ultra--light mass required 
($m \sim 10^{-23}$eV).

\vskip 0.1cm
In the case with $\lambda \neq 0$, the transition between
radiation--like and matter--like behaviors happens even later, just before
equality; therefore, the field density during radiation domination has
to be very close to that of photons and neutrinos. Our simulation
reveals that $\Omega_{\rm cdm} h^2 = 0.13$ today obtains from
$\rho_{\phi} / \rho_{\rm rad} \simeq 0.6$ during radiation domination;
in other words, the field contributes to the number of
relativistic degrees of freedom as an effective neutrino number
$\Delta N_{\rm eff} \simeq 5$, in contradiction with the BBN
constraint $|\Delta N_{\rm eff}| \leq 1$. Moreover, the perturbations
which enter into the Hubble radius before the transition --
which would be probed today by the spectrum of CMB anisotropy and the
linear matter power spectrum, since their comobile wavenumber is
bigger than $k \sim 10^{-2}$~Mpc$^{-1}$ -- would be
suppressed with respect to CDM perturbations (essentially, like for hot
dark matter). These problems were previously noticed by Peebles
\cite{peebles} in the case of a real scalar field. As a possible
solution, Peebles suggests a polynomial self--coupling with a power slightly 
smaller than four. In fact, in order to save this scenario, it would be enough
to increase the redshift of the transition by a factor ten or so.
We conclude that the model with a quartic self--coupling is marginally
excluded by cosmological constraints, but that any small deviations
from the simple framework studied here are worth being investigated.

\vspace{0.5cm}
Our analysis is still incomplete. We said that the conservation of the
charge $J_0 = i (\phi^* \dot{\phi} - \phi \dot{\phi}^*)$ was crucial
for the stability of the condensates. Charge conservation also
provides further constraints between, on the one hand, the
cosmological background of the field in the early Universe, and on the
other hand, the current distribution of dark matter in the form of
field condensates. Also, it can give some hints on possible mechanisms for
the generation of the field density in the early Universe.
In terms of quanta, the conservation of charge implies a constant
number of bosons minus anti-bosons inside a comoving volume:
\beq
Q = (n_{\phi} - n_{\bar{\phi}}) a^3 = {\rm cte}\;\; ,
\eeq 
where ($n_{\phi}$, $n_{\bar{\phi}}$) are the number density of
bosons and of their CP conjugates. We want to
evaluate this number today in the form of galactic dark matter.
A priori, the rotating phase inside each condensate
is fixed up to an arbitrary sign: $\theta(t) \simeq \pm \, m \, t$.
So, some halos can be made of bosons, and some others of
anti--bosons. This may occur, for instance, if the scalar field was
populated during a phase transition: after the transition, the initial field
distribution could be very homogeneous, but with an
arbitrary phase distribution, leading to domains with positive and 
negative values of $\omega=d \theta /dt$. A priori, such a disordered 
initial configuration --
with no phase coherence -- may not exclude the subsequent formation
of Bose condensates.
In this case, the mean number density of
bosons minus anti--bosons today can be arbitrarily close to zero, and
the conservation of charge does not give any new constraint.
Note however that
two halos of opposite charge could annihilate and radiate
out a massive amount of energy. Although this issue would deserve a
more careful study, it is probably in conflict with observations.

\vskip 0.1cm 
Let us consider now the case where all galaxies carry a
charge with same sign. This can occur if the field underwent
inflation -- or was coupled to a field that underwent inflation -- in
such way to be very homogeneous in the early Universe, as assumed in
section III. Then, at the beginning of the matter--like regime, the
phase would be coherent even on super--horizon scales, and all
condensates would form with the same rotating phase. We can estimate
the charge in galaxies, $Q_{\rm gal}$, by multiplying the typical charge 
of a single halo by the number density of halos. This charge must be
smaller than or equal to the total charge of the cosmological 
homogeneous background, $Q_{\rm tot} = \omega \, \sigma^2 a^3$.

\vskip 0.1cm
In the $\lambda=0$ case, the charge
inside one halo is given by \cite{tdlee,AJP_1}: \beq N \sim
\sqrt{\frac{\sigma_0}{M_P}} \, \frac{M_P^2}{m^2} \sim 10^{99} \;.
\eeq Under the very crude assumption that the Universe contains in
average one halo per volume ${\cal V}=1$~Mpc$^3$, we find a mean
density today \beq n_{\phi} = Q_{\rm gal} \; a_0^{-3} 
\sim N/{\cal V} \sim 10^{25}
{\rm cm}^{-3} \; .  \eeq This number is extremely large, $10^{22}$
times bigger than the present number density of photons. This is a
strong indication that the model is not realistic: it would be very
difficult to generate such a huge charge in the early Universe. In
fact we can even completely exclude the model by comparing $Q_{\rm
gal}$ with the total charge of the cosmological background. We saw in
section III, Eqs.(\ref{nospint}, \ref{reallynospint}), that the
existence of a plateau for $\rho_{\phi}$ during radiation domination
imposes today a field dynamics close to that of an oscillating real
axion, rather than spintessence.  For spintessence, the number density
$n_{\phi} = Q_{\rm tot} \, a^{-3}$ is equal to $\rho_{\phi}/m$ as can
be seen from Eq.(\ref{densityFD}) with $E_u=1$. For an oscillating
axion, most of the kinetic energy is in the radial direction and
$n_{\phi} \ll \rho_{\phi}/m$ since the term $E_u$ in
relation~(\ref{densityFD}) is now much larger than 1. Because
we must be in the latter case at least during matter
domination, we find the following upper bound on the total charge
today:
\beq n_{\phi} = Q_{\rm tot} \, a_0^{-3} \ll \frac{\Omega_{\rm cdm} \,
\rho_{\rm c}^0}{m}
\simeq 10^{23} {\rm cm}^{-3} \;.  \eeq We find $Q_{\rm tot} \ll Q_{\rm
gal}$, which is impossible. We
conclude that the present dark matter scenario, based on a complex
free scalar field forming galactic halos after Bose condensation, is
not consistent -- at least when the field is assumed to be homogeneous
in the early Universe, and today all condensates carry a charge with
the same sign. It would certainly be interesting to investigate the opposite
scenario, with a homogeneous initial density but random phases, with
the drawback that halos and ``anti-halos'' may annihilate.

\vskip 0.1cm
As we have seen, the case $\lambda \neq 0$ is already marginally
excluded by cosmological constraints, but it is worth calculating the
various charges also for this model. An individual halo has got a
charge -- see relation~(\ref{relation_34}) --
\beq
|N| \sim \frac{\sigma_0^{2} \, \lambda^{3/2} M_P^{3}}{m^5}
\sim \left( \frac{1 {\rm eV}}{m} \right) 10^{75} \;\; .
\eeq
Under the assumption that all halos have a positive charge, one finds
\beq
n_{\phi} = Q_{\rm gal} \, a_0^{-3} \sim N / {\cal V} \sim
30 \, {\rm cm}^{-3} \;
\left( \frac{1 {\rm eV}}{m} \right) \;\; .
\eeq
We are led to two intriguing coincidences. First, with $\lambda$ of
order one -- and therefore $m$ of order 1~eV -- the field number density
is of the same order of magnitude as that of photons for which
$n_{\gamma} = 400$~cm$^{-3}$. Second, provided that the field behaves
cosmologically as spintessence, with $\rho_{\phi} \sim m~n_{\phi}$,
we can calculate the total charge and find $Q_{\rm tot} \sim Q_{\rm gal}$:
the cosmological and astrophysical charges are
consistent. Therefore, the scenario requires a mechanism in the early
Universe that would fine--tune both $n_{\phi}$ and $\rho_{\phi}$ to
some values very close to $n_{\gamma}$ and $\rho_{\gamma}$.

\vskip 0.1cm
As we already said, this scenario
with a quadratic coupling is plagued by inherent difficulties to
produce small--scale perturbations, and by an inconsistency with the number of
relativistic degrees of freedom indicated by nucleosynthesis. However,
one should retain two positive features. First, any small modification
of the scenario that would shift by a factor ten the redshift of the transition
between radiation--like and matter--like
behaviors would evade these difficulties.
Second, the initial conditions for the field number density and energy density
should be surprisingly close to those of photons.

\vspace{0.5cm}
Throughout this discussion, we tried to give some arguments both in
favor and against the two models considered here, based on two different
scalar potentials. In their present form, none of these
models can survive. However, we believe that one should retain the
many positive indications discussed before as an
encouragement for investigating other variants of scalar field dark
matter. The fact that the two scalar potentials lead to very different
conclusions already shows how rich and unpredictable is the field.

\section*{Acknowledgements}

J.~L. would like to thank J.~Garcia-Bellido, S.~Khlebnikov,
A.~Riotto and M.~Shaposhnikov for illuminating discussions.



%
\end{document}